\documentclass{PoS}
\usepackage{graphicx}
\usepackage{epsfig}
\usepackage{subfigure}
\usepackage{latexsym}
\usepackage{amsmath}
\usepackage{float}

\newcommand{\Tr}{\textrm{Tr}}

\newcommand{\NC}{N_{\rm C}}
\newcommand{\NF}{N_{\rm F}}

\newcommand{\TR}{T_{\rm{R}}}

\newcommand{\csw}{c_{\rm{sw}}}
\renewcommand{\i}{\rm{i}}

\renewcommand{\S}[1]{S_{\rm{#1}}}

\newcommand{\be}{\begin{equation}}
\newcommand{\ee}{\end{equation}}
\newcommand{\bea}{\begin{eqnarray}} 
\newcommand{\eea}{\end{eqnarray}}

\newcommand{\bmp}{\noindent\begin{minipage}{16cm}}
\newcommand{\emp}{\end{minipage}\vskip 7mm} 
\def\lsim{\mathrel{\raise.3ex\hbox{$<$\kern-.75em\lower1ex\hbox{$\sim$}}}}
\def\gsim{\mathrel{\raise.3ex\hbox{$>$\kern-.75em\lower1ex\hbox{$\sim$}}}}

\newcommand{\comment}[1]{}

\title{Effect of the Schr\"odinger functional boundary conditions on the convergence of step scaling}

\ShortTitle{Convergence of step scaling}

\author{\speaker{Tuomas Karavirta}\\
	Department of Physics, P.O.Box 35 (YFL), 
        \\ FI-40014 University of Jyv\"askyl\"a, Finland, 
        \\ and 
  	    \\ Helsinki Institute of Physics, P.O.~Box 64, 
  	    \\ FI-00014 University of Helsinki, Finland\\
	E-mail: \email{tuomas.karavirta@jyu.fi}}

\author{Kari Rummukainen\\
 	Department of Physics and Helsinki Institute of Physics,\\
 	P.O.Box 64, FI-00014 University of Helsinki, Finland\\
	Email: \email{kari.rummukainen@helsinki.fi}}

\author{Kimmo Tuominen\\
	Department of Physics, P.O.Box 35 (YFL), 
        \\ FI-40014 University of Jyv\"askyl\"a, Finland, 
        \\ and 
  	    \\ Helsinki Institute of Physics, P.O.~Box 64, 
  	    \\ FI-00014 University of Helsinki, Finland\\
	Email: \email{kimmo.i.tuominen@jyu.fi}}

\abstract{Recently several lattice collaborations have studied the scale dependence of the coupling in 
theories with different gauge groups and fermion representations using the Schr\"odinger functional 
method. This has motivated us to look at the convergence of the perturbative step scaling to its 
continuum limit with gauge groups SU(2) and SU(3) with Wilson fermions in the fundamental, adjoint 
or sextet representations. We have found that while the improved Wilson action does remove the 
linear terms from the step scaling, the convergence is extremely slow with the standard choices 
of the boundary conditions for the background field. We show that the situation can be improved by 
careful choice of the boundary fields.}

\FullConference{The 30th International Symposium on Lattice Field Theory\\
		 June 24 - 29,  2012\\
		 Cairns, Australia \vspace{1cm}\\
		 HIP-2012-20/TH}

\begin{document}

\section{Introduction and motivation}

\begin{figure}[H]
\centering
\includegraphics[scale=0.18]{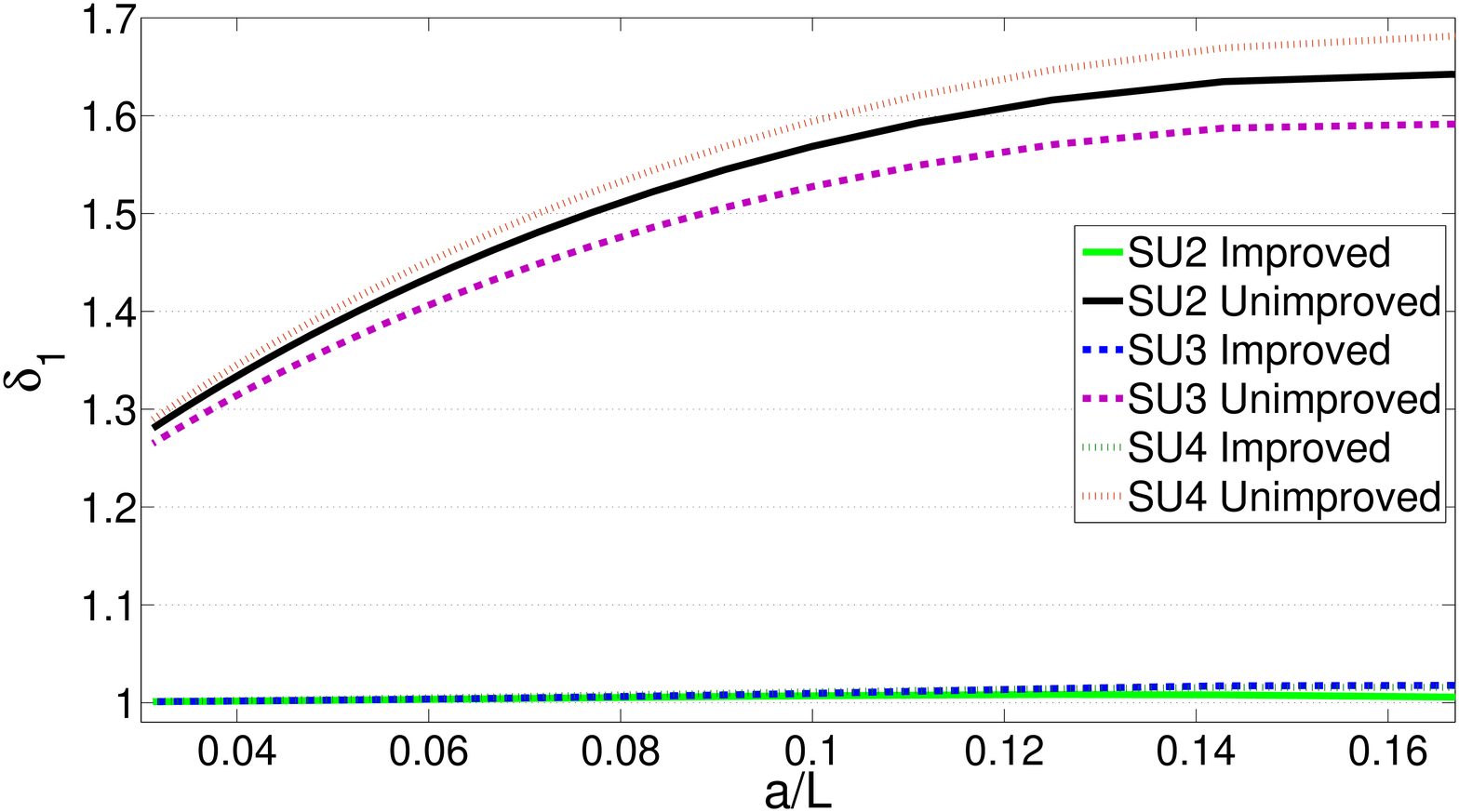}
\includegraphics[scale=0.18]{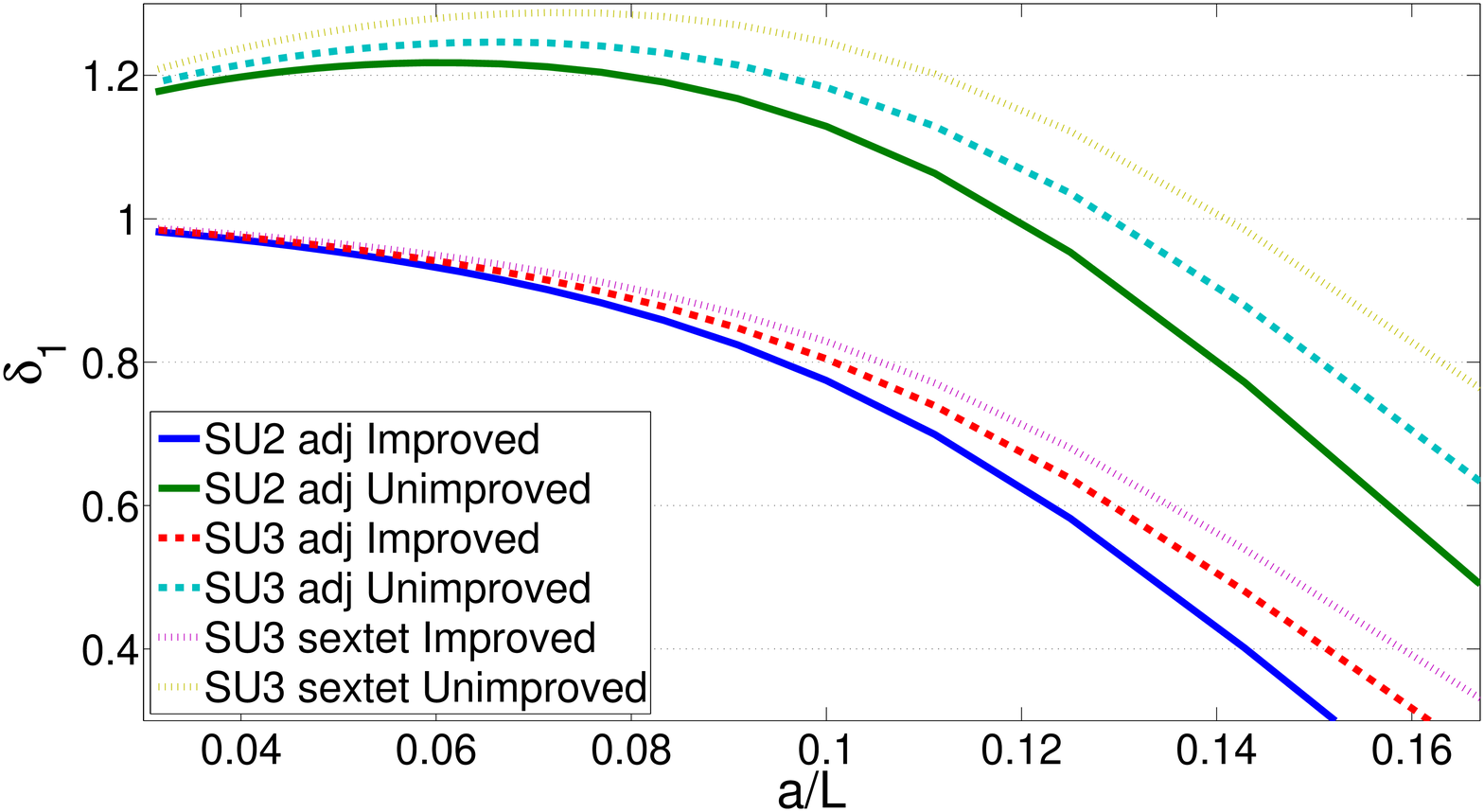}
\caption{Leading perturbative finite volume corrections to the continuum normalized fermion contribution of the lattice step scaling function with standard
Schr\"odinger functional boundary conditions (described below). 
Fundamental fermions on the left panel and higher representation fermions on the right.}
\label{Stepscaling fundamental&higher}
\end{figure}

Measuring the scale evolution of the coupling in gauge theories with matter fields in higher representation has been studied intensively over the past few years \cite{Giedt}. It has been shown in \cite{Sint:1995ch,Sommer:1997jg,Karavirta:2011mv} that for SU(2) and SU(3) with fundamental fermions the lattice step scaling function converges rapidly to its continuum limit when one uses ${\cal{O}}(a)$ improved Wilson-clover action.
This can also easily be seen in the left panel of figure \ref{Stepscaling fundamental&higher}, where the fermionic contribution, normalized to the contunuum value, is shown. However, the situation changes when one considers fermions in the higher representations, which can be seen from the right panel of figure \ref{Stepscaling fundamental&higher}; see also \cite{Sint:2011gv}.

Clearly the fermionic step scaling function for higher representations has large $\mathcal{O}(a^2)$ contributions, which are absent in the step scaling for the fundamental representation fermions. In the following we will study the effect of the Schr\"odinger functional boundary conditions on the convergence of step scaling \cite{Karavirta:2012qd}. 

\section{Theoretical background}
We use the standard Wilson-clover lattice action
\be
S = S_G+S_F+\delta S_{G,b}+S_{gf}+S_{FP},
\label{Eq:aktion}
\ee
with the usual single-plaquette action, $S_G$. The effect of higher representations appears in the Wilson fermion action, $S_F$, where the parallel transporter appearing in the gauge covariant lattice derivatives must be transformed into appropriate representation. 

With perturbative calculations, one must fix the gauge and this is reflected in the action via terms $\S{gf}$ and $\S{FP}$. The specific form of these terms is only important while calculating the contribution of the gauge sector to observables. Since we will mainly be focused on the fermionic contribution, we refer to the original article \cite{Luscher:1992an} for details.

For periodic boundaries, the ${\cal{O}}(a)$ improvement is  obtained by nonperturbatively tuning the coefficient $\csw$ of the Sheikholeslami-Wohlert -term \cite{Sheikholeslami:1985ij,Wohlert:1987rf}, 
\be
\csw\frac{\i a}{4}\sigma_{\mu\nu} F_{\mu\nu}(x),
\ee
which we assume to be contained in the Wilson fermion action $S_F$. For the perturbative analysis
described here we set $\csw=1$. Also, for the fermion field ``twist" in the spatial directions \cite{Sint:1995rb}, $\psi(x+L\hat{k})=e^{\i\theta_k}\psi(x)$, we use $\theta=\pi/5$ throughout.

We wish to apply the Schr\"odinger functional scheme, i.e. introduce a constant background 
field by setting boundary conditions for the gauge fields at times $T=0$ and $T=L$, while retaining periodic boundaries in the spatial directions. This background field introduces 
further  ${\cal{O}}(a)$ contributions which are cancelled by the boundary counterterms 
$\delta S_{G,b}$ in the action (\ref{Eq:aktion}). We will now describe the analysis of their effects in more detail.


The boundary fields used for SU(2) are
\be
C_k=\frac{\i}{L}\left(\begin{array}{cc}\phi_1 & 0 \\ 0 & \phi_2 \end{array}\right), \qquad C'_k=\frac{\i}{L}\left(\begin{array}{cc}\phi'_1 & 0 \\ 0 & \phi'_2 \end{array}\right),\quad k=1,2,3,
\label{CkSU2}
\ee
where
\be
\phi_1=-\eta,\qquad
\phi_2=\eta, \qquad
\phi'_1=\eta-\rho,\qquad
\phi'_2=\rho-\eta.
\label{SU(2)boundary}
\ee
The standard choice for the angles are $\eta=\frac{\pi}{4}$ and $\rho=\pi$ \cite{Luscher:1992zx}. For SU(3) we used
\be
C_k=\frac{\i}{L}\left(\begin{array}{ccc}\phi_1 & 0 & 0\\ 0 & \phi_2 & 0 \\ 0 & 0 & \phi_3  \end{array}\right), \qquad C'_k=\frac{\i}{L}\left(\begin{array}{ccc}\phi'_1 & 0 & 0\\ 0 & \phi'_2 & 0 \\ 0 & 0 & \phi'_3  \end{array}\right),\quad k=1,2,3,
\label{CkSU3}
 \ee
where
\bea
\phi_1&=&\eta-\rho,\qquad\quad
\phi_2=\eta(\nu-\frac{1}{2}),\qquad
\phi_3=-\eta(\nu+\frac{1}{2})+\rho, \\
\phi'_1&=&-\phi_1-4\rho,\,\,\,\,\quad
\phi'_2=-\phi_3+2\rho,\qquad
\phi'_3=-\phi_2+2\rho.
\label{SU(3)boundary}
\eea
The standard choice for the angles are $\eta=0$, $\rho=\frac{\pi}{3}$ and $\nu=0$ \cite{Luscher:1993gh}. 

\comment
{When one uses fermions in a higher representation these boundary fields have to be transformed to the corresponding representation. For the adjoint representation the transformation is 
\begin{equation}
C_k({\rm{adj}})_{ab}=2 \Tr (T^a C_k({\rm{F}})T^b C_k({\rm{F}})^\dagger),
   \label{eq:adjproject}
\end{equation}
where $T^a$ are the generators of the fundamental representation. This works also for $C'_k$. For the (anti)symmetric representation the transformation is 
\begin{equation}
C_k({\rm{(a)s}})_{ab}=\pm \Tr (K^a C_k({\rm{F}})K^b C_k({\rm{F}})),
   \label{eq:asproject}
\end{equation}
where the upper (lower) sign refers to the (anti)symmetric representation and the matrices $K^a$ form a basis in the space of (anti)symmetric $N\times N$ matrices. After the transformation one has to diagonalize the resulting matrix, in order to get the correct boundary field.}

With the boundary matrices $C_k$ and $C'_k$ from \eqref{CkSU2} or \eqref{CkSU3}, 
 we end up with a background field of the form
\be
B_0=0,\qquad B_k=(x_0 C'_k+(L-x_0) C_k)/L,\quad k=1,2,3.
\ee
Using the path integral representation $Z$ of the Schr\"odinger functional
the effective action is
\be
\Gamma= -\ln Z= g_0^{-2} \Gamma_0+\Gamma_1+\mathcal{O}(g_0 ^2).
\ee
The running coupling can be determined by studying how the system reacts to the change 
of the background field. Using the effective action, one defines the running coupling and 
obtains its perturbative expansion as
\bea
\bar{g}^2(L)=\frac{\partial \Gamma_0/ \partial\eta}{\partial \Gamma/ \partial\eta}=\frac{\partial \Gamma_0/ \partial\eta}{\partial (\Gamma_0+\Gamma_1)/ \partial\eta}=g_0 ^2+p_1(L) g_0 ^4+\mathcal{O}(g_0 ^6),
\label{running coupling}
\eea
where 
\be
p_1(L)= -\frac{\partial \Gamma_1/ \partial\eta}{\partial \Gamma_0/ \partial\eta}=p_{1,0}(L)+\NF p_{1,1}(L),
\label{p_1}
\ee
and the quantity $p_1$ has been split into gauge and fermionic parts, $p_{1,0}$ and $p_{1,1}$, 
respectively.

The $\mathcal{O}(a)$ errors arising from the fixed boundary conditions at times $t=0$ and $t=T$   are removed by the countertems \cite{Luscher:1996sc}
\bea
\delta S_{G,b}&=&\frac{1}{2 g_0 ^2}(c_s-1)\sum_{p_s}{\rm{Tr}}[1-U(p_s)]+\frac{1}{g_0 ^2}(c_t-1)\sum_{p_t}{\rm{Tr}}[1-U(p_t)],\\
\delta S_{F,b}&=&a^4 (\tilde{c}_s-1)\sum_{\vec{x}}[\hat{O}_s(\vec{x})+\hat{O}'_s(\vec{x})]+a^4 (\tilde{c}_t-1)\sum_{\vec{x}}[\hat{O}_t(\vec{x})-\hat{O}'_t(\vec{x})].
\eea
For the precise form of the operators $\hat{O}_{s,t}$ and  $\hat{O}_{s,t}$ we refer to 
\cite{Luscher:1996sc}. Together with the bulk improvement coefficient $\csw$, tuning the coefficients 
$c_s,c_t,\tilde{c}_s,\tilde{c}_t$ to their proper values we can remove all $\mathcal{O}(a)$ errors.

For the electric background fields which we consider here, the terms proportional to $c_s$ do not 
contribute. Also, if we set the fermionic fields to zero on the boundaries, the counterterm proportional 
to $\tilde{c}_s$ vanishes. However the two terms proportional to $c_t$ and $\tilde{c}_t$ remain non-
zero. The $\tilde{c}_t$ term corrects the mass of the fermions at times $T=a$ and $T=L-a$ and the 
$c_t$ term changes the weight of the time-like plaquettes on the boundary. In this work we set 
$\tilde{c}_t=1$ and $c_t$ to its one loop perturbative value 
\cite{Sint:1995ch,Karavirta:2011mv,Luscher:1992an,Luscher:1993gh}. 
This gives us one loop $\mathcal{O}(a)$ improvement to the lattice step scaling function.

The step scaling function and its perturbative expansion to one loop are
\begin{eqnarray}
\Sigma(u,s,L/a)&=&g^2(g_0,sL/a)\vert_{g^2(g_0,L/a)=u},\\
&=&u+\left[\Sigma_{1,0}(s,L/a)+\Sigma_{1,1}(s,L/a) \NF\right] u^2+\mathcal{O}(u^3).
\label{eq:Step scaling}
\end{eqnarray}
Using the perturbative expansion of the coupling from \eqref{running coupling} in the perturbative 
formula for the step scaling function, we obtain for the one loop coefficients the following useful 
formulas: 
\bea
\Sigma_{1,0}(s,L/a)&=&p_{1,0}(sL)-p_{1,0}(L),\\
\Sigma_{1,1}(s,L/a)&=&p_{1,1}(sL)-p_{1,1}(L).
\eea
We also introduce the variable 
\begin{equation}
\delta_i=\frac{\Sigma_{1,i}(2,L/a)}{\sigma_{1,i}(2)}=\frac{\Sigma_{1,i}(2,L/a)}{2b_{0,i}\ln 2},\qquad i=0,1,
\label{delta_i}
\end{equation}
which is the ratio of the perturbative step scaling and its continuum limit. This variable is useful in illustrating the convergence of the step scaling. In equation \eqref{delta_i} we used
\begin{equation}
b_{0,0}=11\NC/(48\pi^2),\qquad b_{0,1}=-\TR/(12\pi^2),
\end{equation}
which are the one loop coefficients of the perturbative beta function.

The choice of the boundary fields in \eqref{SU(2)boundary} and \eqref{SU(3)boundary} is in no way unique. In fact one can choose the form of the boundary fields and the values of the angles $\eta$, $\rho$ and $\nu$ quite freely. The only limitation is that the fields $\phi$ and $\phi'$ have to belong to the so called fundamental domain. This consists of all the boundary fields that satisfy the equations
\begin{equation}
\phi_1< \phi_2 < \ldots <\phi_n,\qquad\quad |\phi_{i}-\phi_{j}|<2\pi, \text{for all }i,j,\qquad\quad \sum_{i=1} ^N \phi_i=0.
\label{fundamental domain}
\end{equation}
Boundary fields of this type lead to a unique (up to a gauge transformation) minimal action, as has been shown in \cite{Luscher:1992an}. 

We take the fields $\phi$ and $\phi'$ that were introduced in equations \eqref{SU(2)boundary} and \eqref{SU(3)boundary} and choose variables $\eta$ and $\rho$ to be free parameters. The equation \eqref{fundamental domain} then gives us the range of allowed values for $\eta$ and $\rho$. 

\section{Numerical results}
Here we will present absolute errors of the the step scaling functions $\Sigma_{1,1}(L=10,s=2)$ as a function of $\eta$ and $\rho$ for the adjoint representations of SU(2) and SU(3) and the sextet representation of SU(3). The plotted variable thus is $|\delta_{1,1}-1|$.

\begin{figure}[H]
\centering
\includegraphics[height=5cm,width=12cm]{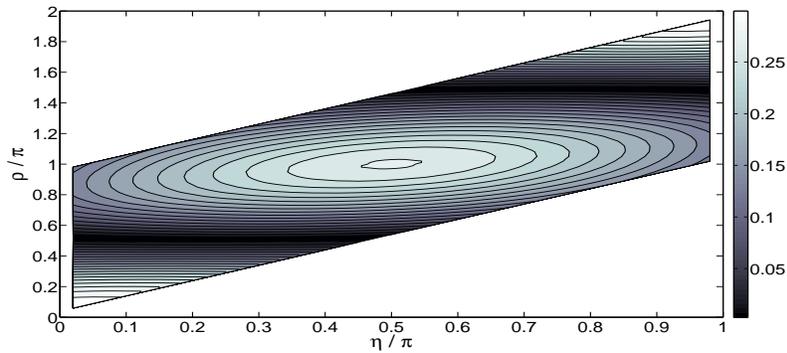}
\caption{Fermionic part of the Lattice step scaling function for SU(2) with adjoint fermions at $L=10$. Optimal choice is $\rho=\frac{\pi}{2}$ and $\eta=\frac{\pi}{8}$.}
\label{SU2_adj_contour}
\end{figure}

Consider first SU(2) with two adjoint Dirac fermions. There are two darker regions in the figure \ref{SU2_adj_contour}, indicating the areas where the discretization errors are the smallest. These two values of $\rho$, $\pi/2$ and $3\pi/2$ are actually equivalent. The $\eta$ dependence is weak, and $\eta$ can be selected from the values within the fundamental domain quite freely (this turns 
out to be true also for SU(3)). However the value $\eta=\rho/2$ must be excluded, since the step scaling function diverges at that point. 

\begin{figure}[H]
\centering
\includegraphics[height=5cm, width=12cm]{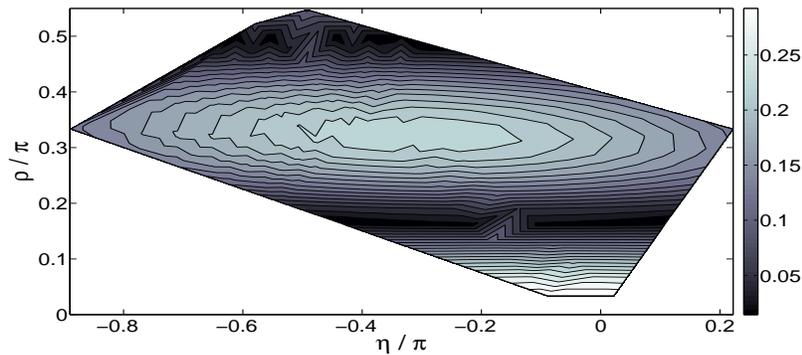}
\caption{Fermionic part of the Lattice step scaling function for SU(3) with adjoint fermions at $L=10$.  Optimal choice is $\rho=\frac{\pi}{6}$ and $\eta=-\frac{\pi}{9}$.}
\label{SU3_adj_contour}
\end{figure}

\begin{figure}[H]
\centering
\includegraphics[height=5cm, width=12cm]{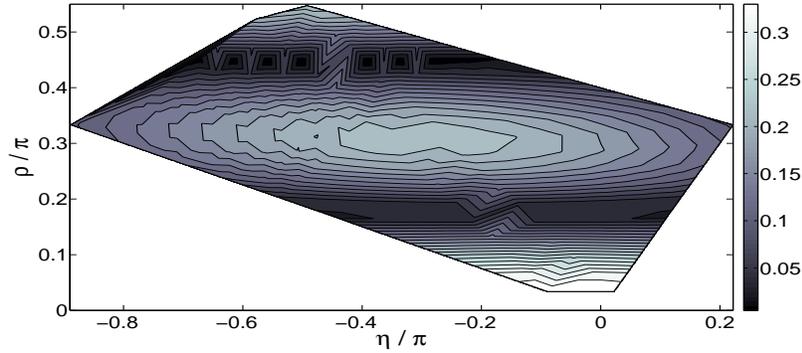}
\caption{Fermionic part of the Lattice step scaling function for SU(3) with sextet fermions at $L=10$. Optimal choice is $\rho=\frac{67\pi}{150}$ and $\eta=-\frac{\pi}{3}$.}
\label{SU3_sextet_contour}
\end{figure}

Then turn to SU(3) with two Dirac fermions either in the adjoint or the sextet representation. Figure \ref{SU3_adj_contour} shows also two darker areas for SU(3) with adjoint fermions. More detailed analysis shows, that in the $\rho=\pi/6$ area the $\mathcal{O}(a^2)$ effects are smaller. In figure \ref{SU3_sextet_contour}, which shows the step scaling function for SU(3) sextet, there is only one dark region at $\rho=67\pi/150$. We also found out that, these optimal values of $\eta$ and $\rho$ are independent of $L$. Change in $L$ only modifies the scale of the errors.

\begin{figure}[H]
\centering
\includegraphics[scale=0.25]{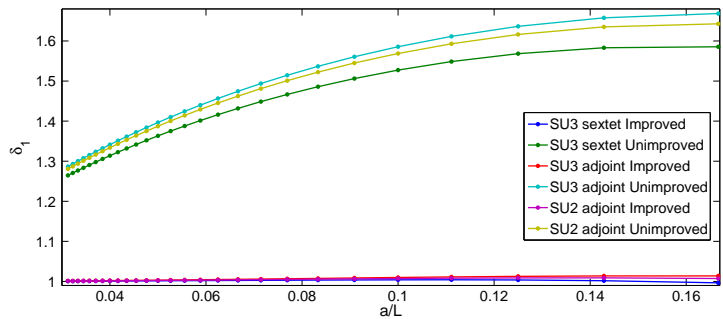}
\caption{Fermionic part of the Lattice step scaling function for higher representation fermions with the new boundary conditions.}
\end{figure}

We also want to be sure that the value of the step scaling parameter $s$ does not have an effect on the convergence of step scaling. While the convergence does improve slightly as $s$ is increased, the change is still quite small and not enough to cure the higher order lattice effects present in the step scaling functions with the old boundary conditions\footnote{From now on we will use the terms "old" and "new" boundary conditions. The old boundary conditions refer to the conditions presented in equations \eqref{SU(2)boundary} and \eqref{SU(3)boundary} with the standard choice of the parameters $\rho$ and $\eta$. With the term new boundary conditions we refer to the optimal choice of the parameters $\rho$ and $\eta$ for a specific symmetry group and fermion representation as stated in this section.}. We have also checked that $s$ has no effect on the preferred values of $\eta$ and $\rho$.

One has to keep in mind that changing the boundary conditions could in principle also affect the convergence of the gauge part of the step scaling function. This happens to be true for SU(2), but fortunately the effects are roughly a hundred times smaller. Thus one can freely choose the boundary conditions that are optimal for the fermionic part without compromising the convergence of the step scaling function. In fact the convergence is even faster if one uses the new boundary conditions that are optimized for the adjoint fermions. We expect similar behavior from the gauge sector of the other gauge groups. 

\section{Conclusions}
We have provided a cure to the slow convergence of the lattice step scaling function with higher 
representation fermions. Our method has the drawback, that in simulations with higher 
representation fermions and new boundary condition  the signal for measuring the scale evolution of 
the coupling constant  can be smaller than it is with the old boundary conditions. 
However one has to get rid of the higher order lattice artifacts, to get reliable results. Other solutions to 
this problem exist for SU(3) \cite{SintVilaseca}. 
A combination of these methods might provide a way to get the 
smallest possible lattice artifacts while retaining also the signal optimally high. 

\acknowledgments
T.K. is supported by the Magnus Ehrnrooth foundation and by University of Jyv\"askyl\"a Faculty of Mathematics and Science. K.R. has been supported by the Academy of Finland grant 1134018.


\begin{thebibliography}{99}

\bibitem{Giedt}
J.\, Giedt, in these proceedings.

\bibitem{Sint:1995ch}
  S.~Sint and R.~Sommer,
  Nucl.\ Phys.\  B {\bf 465}, 71 (1996)
  [arXiv:hep-lat/9508012].
  
\bibitem{Sommer:1997jg}
  R.~Sommer,
  Nucl.\ Phys.\ Proc.\ Suppl.\  {\bf 60A}, 279 (1998)
  [arXiv:hep-lat/9705026].
  
\bibitem{Karavirta:2011mv} 
  T.~Karavirta, A.~Mykkanen, J.~Rantaharju, K.~Rummukainen and K.~Tuominen,
  JHEP {\bf 1106}, 061 (2011)
  [arXiv:1101.0154 [hep-lat]].
  
\bibitem{Sint:2011gv} 
  S.~Sint and P.~Vilaseca,
  PoS LATTICE {\bf 2011}, 091 (2011)
  [arXiv:1111.2227 [hep-lat]];
 
  
\bibitem{Karavirta:2012qd} 
  T.~Karavirta, K.~Tuominen and K.~Rummukainen,
  Phys.\ Rev.\ D {\bf 85}, 054506 (2012)
  [arXiv:1201.1883 [hep-lat]].
  
\bibitem{Luscher:1992an}
  M.~Luscher, R.~Narayanan, P.~Weisz and U.~Wolff,
  Nucl.\ Phys.\  B {\bf 384}, 168 (1992)
  [arXiv:hep-lat/9207009].

  
\bibitem{Sheikholeslami:1985ij}
  B.~Sheikholeslami and R.~Wohlert,
  Nucl.\ Phys.\  B {\bf 259}, 572 (1985).



\bibitem{Wohlert:1987rf}
  R.~Wohlert, {\em Improved Continuum Limit Lattice Action For Quarks},
  DESY87/069
  
\bibitem{Sint:1995rb} 
  S.~Sint,
  Nucl.\ Phys.\ B {\bf 451}, 416 (1995)
  [hep-lat/9504005].


  
\bibitem{Luscher:1992zx}
  M.~Luscher, R.~Sommer, U.~Wolff and P.~Weisz,
  {\em Computation Of The Running Coupling In The SU(2) Yang-Mills Theory,}
  Nucl.\ Phys.\  B {\bf 389}, 247 (1993)
  [arXiv:hep-lat/9207010].
  
  
\bibitem{Luscher:1993gh}
  M.~Luscher, R.~Sommer, P.~Weisz and U.~Wolff,
  Nucl.\ Phys.\  B {\bf 413}, 481 (1994)
  [arXiv:hep-lat/9309005].
  
\bibitem{Luscher:1996sc}
  M.~Luscher, S.~Sint, R.~Sommer and P.~Weisz,
  Nucl.\ Phys.\  B {\bf 478}, 365 (1996)
  [arXiv:hep-lat/9605038].

\bibitem{SintVilaseca}
S. Sint and P. Vilaseca, in these proceedings.

\end{thebibliography}
\end{document}